\documentclass[aps,prd]{revtex4}

\usepackage{amsmath}
\usepackage{amssymb}
\usepackage{bm}
\usepackage{graphicx}
\usepackage{multirow}

\usepackage{textcomp}

\allowdisplaybreaks[1]

\begin{document}

\title{Dispersive determination of neutrino mass ordering}
\author{Hsiang-nan Li}
\affiliation{Institute of Physics, Academia Sinica,
Taipei, Taiwan 115, Republic of China}

\date{\today}

\begin{abstract}
We argue that the mixing phenomenon of a neutral meson formed by a fictitious massive quark  
will disappear, if the electroweak symmetry of the Standard Model (SM) is restored at a 
high energy scale. This disappearance is taken as the high-energy input for the dispersion 
relation, which must be obeyed by the width difference between two meson mass eigenstates. 
The solution to the dispersion relation at low energy, i.e., in the symmetry broken phase, 
then connects the Cabibbo-Kobayashi-Maskawa (CKM) matrix elements to the quark masses 
involved in the box diagrams responsible for meson mixing. It is demonstrated via the 
analysis of the $D$ meson mixing that the typical $d$, $s$ and $b$ quark masses demand the 
CKM matrix elements in agreement with measured values. In particular, the known numerical 
relation $V_{us}\approx \sqrt{m_s/m_b}$ with the $s$ ($b$) quark mass $m_s$ ($m_b$) can be 
derived analytically from our solution. Next we apply the same formalism to the mixing of 
the $\mu^- e^+$ and $\mu^+ e^-$ states through similar box diagrams with intermediate 
neutrino channels. It is shown that the neutrino masses in the normal hierarchy (NH), 
instead of in the inverted hierarchy or quasi-degenerate spectrum, match the observed 
Pontecorvo-Maki-Nakagawa-Sakata matrix elements. The lepton mixing angles larger than the 
quark ones are explained by means of the inequality $m_2^2/m_3^2\gg m_s^2/m_b^2$, 
$m_{2,3}$ being the neutrino masses in the NH. At last, the solution for the 
$\tau^-e^+$-$\tau^+e^-$ mixing specifies the mixing angle $\theta_{23}\approx 45^\circ$,
leading to the $\mu$-$\tau$ reflection symmetry. Our work suggests that the fermion masses 
and mixing parameters are constrained dynamically, and the neutrino mass orderings can be 
discriminated by the internal consistency of the SM. 

\end{abstract}


\maketitle

\section{INTRODUCTION}

It has been believed that the parameters in the Standard Model (SM), such as particle masses 
and mixing angles, are free, and have to be determined experimentally. These parameters 
originate from the independent elements of the Yukawa matrices, which cannot be completely 
constrained by symmetries \cite{Santamaria:1993ah}, as the electroweak symmetry is broken. 
Any attempt to explain their values relies on an underlying new physics theory existent at 
high energy, whose low-energy behavior fixes the SM parameters. However, the above 
observation is made at the Lagrangian level without taking into account subtle dynamical 
interplay among the involved gauge and scalar sectors. We have pointed out in recent publications 
\cite{Li:2023dqi,Li:2023yay} that dispersion relations, which physical observables must obey 
owing to analyticity, connect various interactions at different scales, and thus impose stringent 
constraints on the SM parameters. The SM parameters should satisfy dispersive constraints from 
all physical observables in principle. We have demonstrated, by considering those  
which provide efficient constraints \cite{Li:2023dqi,Li:2023yay}, that at least some of the 
SM parameters can be determined dynamically within the model itself.


A dispersion relation links the high- and low-energy properties of an observable, which is 
defined by a correlation function. The high-energy property, calculated perturbatively 
from the correlation function, is treated as an input. The low-energy property is then solved 
directly from the dispersion relation with the given input, which demands specific values for 
relevant particle masses in agreement with measured ones. We have analyzed heavy meson decay 
widths \cite{Li:2023dqi}, written as absorptive pieces of hadronic matrix elements of  
four-quark effective operators, in this inverse-problem approach 
\cite{Li:2020xrz,Li:2020fiz,Li:2020ejs,Xiong:2022uwj}. Starting 
with massless final-state $u$ and $d$ quarks, we found that the solution for the decay 
$c\to du\bar d$ ($b\to c\bar ud$) with heavy-quark-expansion (HQE) inputs leads to the $c$ 
($b$) quark mass $m_c=1.35$ ($m_b= 4.0$) GeV. The requirement that the dispersion relation for 
the $c\to su\bar d$ ($c\to d\mu^+\nu_\mu$, $b\to u\tau^-\bar\nu_\tau$) decay yields the same 
heavy quark mass fixes the strange quark (muon, $\tau$ lepton) mass $m_s= 0.12$ GeV 
($m_\mu=0.11$ GeV, $m_\tau= 2.0$ GeV). The investigation on the dispersion relation 
respected by the correlation function of two $b$-quark scalar (vector) currents, with the input 
from the perturbative evaluation of the $b$ quark loop, returns the Higgs ($Z$) boson mass 114 
(90.8) GeV \cite{Li:2023yay}. A reasoning for why the analyticity can constrain the
SM parameters through dispersion relations for physical observables has been provided in 
\cite{Li:2023yay}.

The successful explanation of the particle masses from 0.1 GeV up to the electroweak scale 
by means of the internal consistency of SM dynamics encourages us to address the fermion 
mixing in the same formalism. We first argue that the mixing phenomenon of a neutral meson 
formed by a fictitious massive quark will disappear, if the electroweak symmetry of the SM 
is restored at a high energy scale \cite{Chien:2018ohd,Huang:2020iya}: internal particles
involved in the box diagrams responsible for the mixing become massless in the symmetric
phase, so the contributions from all intermediate channels cancel simply owing to the 
Glashow–Iliopoulos–Maiani mechanism \cite{Glashow:1970gm}. There are new physics models in 
the literature, which provide a suitable framework for our discussion based on the 
restoration of the electroweak symmetry. For instance, the composite Higgs model proposed 
in \cite{Kaplan:1983fs} meets the purpose well. The electroweak group in their model is 
broken at a scale much lower than the condensate scale, implying the existence of a symmetry
restoration scale which we refer to. The disappearance of the mixing phenomenon then takes 
place in the region above the restoration scale and below the condensate scale. A concrete 
example with the incorporation of a sequential fourth generation, that realizes the 
restoration of the electroweak symmetry at a high energy, can be located in \cite{Li:2025bon}. 
We do not intend to elaborate the detail of this model here, but to establish the dispersive 
constraints that some SM parameters should satisfy, if the electroweak symmetry is restored. 


The restoration of the electroweak symmetry is the only assumption required in our analysis.
The disappearance of the mixing phenomenon
is taken as the high-energy input for the dispersion relation satisfied by the width difference 
between two meson mass eigenstates. The solution to the dispersion relation at low energy, i.e., 
in the symmetry broken phase, effectively binds the Cabibbo-Kobayashi-Maskawa (CKM) matrix 
elements and the quark masses appearing in the box diagrams. Compared to conventional
applications of dispersion relations to processes involving strong interaction, the symmetric
phase in the present setup corresponds to the perturbative region, where high-energy 
dynamics is treated as inputs, and the broken phase corresponds to the nonperturbative
region, where low-energy behaviors are constrained.
It will be elaborated how the typical $d$, $s$ and $b$ quark masses constrain the 
CKM matrix elements through the dispersive relation for the $D$ 
meson mixing. The connection between the fermion flavor structure and the pattern of the Yukawa 
matrices, together with plausible relations among the quark masses and the mixing angles, have 
been speculated \cite{HF78,CS87}. For a recent reference in this direction based on Yukawa matrix 
textures, see \cite{Belfatto:2023qca}. We will derive the known empirical relation 
$V_{us}\approx \sqrt{m_s/m_b}$ \cite{Belfatto:2023qca} with the $s$ ($b$) quark mass $m_s$ 
($m_b$) analytically from our solution. Namely, our work realizes the speculation in the 
literature, and suggests that its underlying theory is the SM itself.


We then perform the similar analysis of the mixing between the $\mu^- e^+$ and 
$\mu^+ e^-$ states, which occurs via the box diagrams containing intermediate neutrino channels. 
The formulas are exactly the same as of the $D$ meson, i.e., $c\bar u$-$\bar cu$ 
mixing, with the quark masses $m_{d,s,b}$ being replaced by the neutrino masses 
$m_{1,2,3}$, and the CKM matrix elements by the Pontecorvo–Maki–Nakagawa–Sakata (PMNS) ones.
It will be shown that the neutrino masses in the normal hierarchy (NH), instead of 
in the inverted hierarchy (IH) or quasi-degenerate (QD) spectrum, match the observed PMNS matrix 
elements. The neutrino mass ordering, whose various scenarios have not been discriminated 
experimentally, has remained as an unsettled issue in neutrino physics \cite{PDG}. Our study 
provides a solid theoretical support for the NH spectrum in the viewpoint of the internal 
consistency of SM dynamics. The neutrino mixing angles larger than the quark ones are then 
accounted for naturally by the inequality $m_2^2/m_3^2\gg m_s^2/m_b^2$ for $m_{2,3}$ 
in the NH. We further examine the $\tau^-e^+$-$\tau^+e^-$ mixing, and find that its
solution requests the angle $\theta_{23}\approx 45^\circ$ in accordance with its 
measured value around the maximal mixing, explaining the $\mu$-$\tau$ reflection symmetry 
\cite{Harrison:2002et}. It is emphasized that the above relations 
between the fermion masses and mixing angles are established without resorting to specific
new ingredients beyond the SM (for recent endeavors on this topic, refer to 
\cite{Alvarado:2020lcz,Xu:2023kfi,Patel:2023qtw,Bora:2023teg,Thapa:2023fxu}).

\section{FORMALISM}


Consider the mixing mechanism of a neutral meson formed by a fictitious massive quark $Q$ 
in the SM. Precisely, we work on the mixing 
between the $Q_L\bar q_L$ and $\bar Q_Lq_L$ states, where $q$ is a light quark and the 
subscript $L$ denotes the left handedness. 
The external states $Q_L\bar q_L$ and $\bar Q_Lq_L$ are formed by the quarks in the 
broken phase, since they provide a large mass scale. The first emissions are thus 
composed of real neutral scalars or weak gauge bosons\cite{Li:2024awx}, and only two 
$W$ boson are exchanged at leading order to induce the mixing. All internal particles 
become massless in the symmetric phase as the external states are heavy enough, and their 
contributions cancel owing to the unitarity of the CKM matrix. The vanishing of a mixing 
observable at high energy will be taken as an input in the dispersive analysis below.
We mention an alternative setup for the same purpose, $Q_L\bar q_L$ scattering into 
$\bar Q_Lq_L$ at arbitrary center-of-mass energy $E$. As $E$ is high enough, all 
intermediate particles become massless, and the corresponding amplitude diminishes.

The dispersion relation for neutral meson mixing is quoted as \cite{Li:2022jxc}
\begin{eqnarray}
M_{12}(s)=\frac{1}{2\pi}\int ds'\frac{\Gamma_{12}(s')}{s-s'},\label{dis}
\end{eqnarray}
where $s$ is the mass squared of the quark $Q$ \cite{Li:2020xrz}, and the application of 
the principal-value prescription to the right-hand side is implicit. The proposed
contour and the location of $s$ have been described in Fig.~1 of Ref.~\cite{Li:2022jxc}.
In the above expression $M_{12}(s)$ and $\Gamma_{12}(s)$ represent the real and imaginary 
pieces of the box-diagram contribution, respectively, which governs the time evolution of 
the fictitious neutral meson. Their analytical properties can be inferred from the explicit 
expressions in Appendix A of Ref.~\cite{BSS}. Simply speaking, the box-diagram contribution 
has no poles but contains branch cuts along the real axis with the thresholds being 
specified below. The piece $\Gamma_{12}$ is related to the width difference between the two 
meson mass eigenstates. As $s$, i.e., the scale involved in the box diagrams is large enough, 
the argument about the disappearance of the mixing phenomenon in the symmetric phase 
implies $M_{12}(s)\approx 0$. For a similar reason, the upper bound of the integration variable
$s'$ in Eq.~(\ref{dis}) can be set to the electroweak symmetry restoration scale. 


The dispersive integral on the right-hand side of Eq.~(\ref{dis}) receives the low-mass 
contribution from $\Gamma_{12}$, which depends on the CKM matrix elements associated with 
various massive intermediate quarks in the symmetry broken phase. 
It has been illustrated that the physical $\Gamma_{12}$ with hadronic thresholds and the
perturbative $\Gamma_{12}$ from the box diagrams with quark-level thresholds give
the same dispersive integral \cite{Li:2022jxc}. This equality has allowed us to solve for
the physical $\Gamma_{12}$ from the dispersion relation, which accommodates the observed 
large $D$ meson mixing parameters. Here we adopt the perturbative $\Gamma_{12}$ for the
evaluation of the dispersive integral in Eq.~(\ref{dis}). The box diagrams
generate the $(V-A)(V-A)$ and $(S-P)(S-P)$ effective operators, which should be handled 
independently. We concentrate on the former contribution, which is expressed as \cite{BSS,Cheng}
\begin{eqnarray}
\Gamma_{12}(s)&\propto&\sum_{i,j}\lambda_{i}\lambda_{j} \Gamma_{ij}(s),\nonumber\\
\Gamma_{ij}(s)&=&
\frac{1}{s^2}
\frac{\sqrt{s^2-2s(m_i^2+m_j^2)+(m_i^2-m_j^2)^2}}{(m_W^2-m_i^2)(m_W^2-m_j^2)}\nonumber\\
& &\times\left\{\left(m_W^4+\frac{m_i^2m_j^2}{4}\right)
[2s^2-4s(m_i^2+m_j^2)+2(m_i^2-m_j^2)^2]+3m_W^2s(m_i^2+m_j^2)(m_i^2+m_j^2-s)\right\},
\label{aij}
\end{eqnarray} 
with the $W$ boson mass $m_W$ and the intermediate quark masses $m_i$ and $m_j$. The
overall coefficient, including the bag parameter and the meson decay constant which 
are irrelevant to the reasoning below, has been suppressed.
For the $D$ meson mixing, $i,j=d,s,b$ label the down-type quarks, and 
$\lambda_i\equiv V^*_{ci}V_{ui}$ are the products of the CKM matrix elements. 
It will be verified that the same conclusion is drawn, when the analysis is performed
based on the perturbative contribution from the $(S-P)(S-P)$ operator.


It seems that the $W$ bosons in the box diagrams for the $c\bar u$-$\bar cu$ mixing can be 
integrated out. First, Eq.~(\ref{aij}) has appeared in the
dispersive determination of the top quark mass from the $t\bar u$-$\bar tu$ mixing 
\cite{Li:2023yay}. The implication on the present subject from this mixing will be discussed
in Sec.~IV, for which the $W$ boson fields should not be integrated out.
Moreover, the box-diagram contribution at
high $s$ is crucial for deriving the constraints on fermion masses and mixing 
angles as seen shortly, in which heavy gauge bosons ought to remain dynamical. It is thus
appropriate to quote the formulas in \cite{BSS,Cheng} directly with the $m_W$ dependence being kept. 
Note that in our previous study on heavy meson decay widths \cite{Li:2023dqi}, we focused on the 
explanation of the heavy quark masses below $m_W$, the region which prefers the 
employment of the effective theory with $W$ boson fields being integrated out.

To diminish the dispersive integral in Eq.~(\ref{dis}) for large $s$, some 
conditions must be met by the CKM matrix elements. We begin with the asymptotic behavior of 
Eq.~(\ref{aij})
\begin{eqnarray}
\Gamma_{ij}(s')\approx \Gamma^{(1)}_{ij}s'+\Gamma^{(0)}_{ij}+\frac{\Gamma^{(-1)}_{ij}}{s'}
+\cdots,\label{exp}
\end{eqnarray}
with the coefficients
\begin{eqnarray}
\Gamma^{(1)}_{ij}&=&\frac{4m_W^4-6m_W^2(m_i^2+m_j^2)+m_i^2m_j^2}
{2(m_W^2-m_i^2)(m_W^2-m_j^2)},\nonumber\\
\Gamma^{(0)}_{ij}&=&-\frac{3(m_i^2+m_j^2)\left[4m_W^4-4m_W^2(m_i^2+m_j^2)+m_i^2m_j^2\right]}
{2(m_W^2-m_i^2)(m_W^2-m_j^2)},\nonumber\\
\Gamma^{(-1)}_{ij}&=&\frac{3(m_i^4+m_j^4)\left[4m_W^4-2m_W^2(m_i^2+m_j^2)+m_i^2m_j^2\right]}
{2(m_W^2-m_i^2)(m_W^2-m_j^2)}.\label{mij}
\end{eqnarray}
Each term $\Gamma_{ij}^{(m)}$ gives contributions scaling like $\Lambda^2/s$, 
$(m_i^2+m_j^2)\Lambda/s$, and $(m_i^4+m_j^4)\ln\Lambda/s$ for $m=1,0$, and $-1$, respectively, 
to the dispersive integral in Eq.~(\ref{dis}), where the variable $\Lambda$ is of order of  
the restoration scale. Suppression on these contributions characterized by large $\Lambda$ is 
necessary for making finite the dispersive integral, which can be achieved only by imposing
\begin{eqnarray}
\sum_{i,j}\lambda_{i}\lambda_{j} \Gamma_{ij}^{(m)}\approx 0,\;\;\;\;m=1,0,-1,\label{cons}
\end{eqnarray}
in view of the variability of $\Lambda$. Because $\Lambda$ is not infinite, the left-hand 
sides of the above relations need not vanish exactly, but to be tiny enough.

Once the conditions in Eq.~(\ref{cons}) are fulfilled, we recast the dispersive integral into
\begin{eqnarray}
\int ds'\frac{\Gamma_{12}(s')}{s-s'}\approx
\frac{1}{s}\sum_{i,j}\lambda_{i}\lambda_{j}g_{ij},\label{cong}
\end{eqnarray}
with the factors
\begin{eqnarray}
g_{ij}\equiv\int_{t_{ij}}^\infty ds'\left[\Gamma_{ij}(s')-
\Gamma_{ij}^{(1)}s'-\Gamma^{(0)}_{ij}-\frac{\Gamma^{(-1)}_{ij}}{s'}\right],\label{gi}
\end{eqnarray}
and the thresholds $t_{ij}=(m_i+m_j)^2$. The approximation $1/(s-s')\approx 1/s$ has been 
applied, which holds well for large $s$, since the integral receives contributions only 
from finite $s'$. The integrand in the square brackets decreases like $1/s^{\prime 2}$, so the 
upper bound of $s'$ in Eq.~(\ref{gi}) can be extended to infinity safely. We place the final 
condition
\begin{eqnarray}
\sum_{i,j}\lambda_{i}\lambda_{j}g_{ij}\approx 0,\label{gij}
\end{eqnarray}
to ensure the almost nil dispersive integral. That is, the realization of Eqs.~(\ref{cons}) 
and (\ref{gij}) establishes a solution to the dispersion relation in Eq.~(\ref{dis})  
at large $s$ with $M_{12}(s)\approx 0$. 

Some remarks are in order. One may wonder whether the divergent pieces in the dispersive integral 
in Eq.~(\ref{dis}) can be removed by subtraction. As the subtraction is implemented by 
introducing one power of $s'-s_1$ into the denominator of the dispersive integral, a subtraction 
constant $M_{12}(s_1)$ appears,
\begin{eqnarray}
\frac{M_{12}(s)-M_{12}(s_1)}{s_1-s}=
\frac{1}{2\pi}\int ds'\frac{\Gamma_{12}(s')}{(s-s')(s_1-s')}.\label{di1}
\end{eqnarray}
If $s_1$ is above $\Lambda$, like $s$, we will have 
$M_{12}(s)\approx 0$, $M_{12}(s_1)\approx 0$ and $(s-s')(s_1-s')\approx ss_1$. The above 
expression then reduces to $\int ds'\Gamma_{12}(s')\approx 0$, which produces the conditions 
identical to Eqs.~(\ref{cons}) and (\ref{gij}). If $s_1$ is below $\Lambda$, we will have 
$M_{12}(s)\approx 0$, $s_1-s\approx -s$ and $s-s'\approx s$. Equation~(\ref{di1}) thus turns
into the same form as Eq.~(\ref{dis}) that we started with. It is more difficult to extract 
constraints from the dispersion relation with $s_1<\Lambda$, and solving such an integral 
equation is not our strategy. Namely, implementing the subtraction either leaves our formalism 
unchanged or complicates the analysis, and does not really tame the divergent behavior of the 
dispersive integral. Note that we stick to the leading-order accuracy in this work as the first 
attempt to extract the relations among fermion masses and mixing angles by means of analyticity. 
Higher-order QCD and electroweak corrections, such as those from the channels involving quark 
pairs through $W$ boson decays, can be taken into account systematically in the future.

\section{QUARK MASSES AND THE CKM MATRIX}

It is apparent that Eqs.~(\ref{cons}) and (\ref{gij}) enforce the connections between the CKM 
matrix elements and the quark masses speculated in the literature, which will be confronted by
the data below. With the unitarity of the CKM matrix, we rewrite these conditions 
for the $D$ meson, i.e., $c\bar u$-$\bar cu$ mixing as
\begin{eqnarray}
& &r^2R_{dd}^{(m)}+2rR_{ds}^{(m)}+1\approx 0,\;\;\;\;m=1,0,-1,i\label{co1}
\end{eqnarray}
with the ratios
\begin{eqnarray}
R_{dd}^{(m)}=\frac{\Gamma_{dd}^{(m)}-2\Gamma_{db}^{(m)}+\Gamma_{bb}^{(m)}}
{\Gamma_{ss}^{(m)}-2\Gamma_{sb}^{(m)}+\Gamma_{bb}^{(m)}},\;\;\;\;
R_{ds}^{(m)}=\frac{\Gamma_{ds}^{(m)}-\Gamma_{db}^{(m)}-\Gamma_{sb}^{(m)}+\Gamma_{bb}^{(m)}}
{\Gamma_{ss}^{(m)}-2\Gamma_{sb}^{(m)}+\Gamma_{bb}^{(m)}},\label{Rij}
\end{eqnarray}
for $m=1,0,-1$. The expression for $m=i$ is similar with $g_{ij}$ being substituted for 
$\Gamma_{ij}^{(m)}$ in Eq.~(\ref{Rij}). We will encounter real $W$-boson production as the 
massive quark mass exceeds the thresholds $m_W+m_i$, $m_W+m_j$ and $2m_W$, whose effects are 
not taken into account in Eq.~(\ref{aij}).
However, these thresholds, much greater than the other scales in the box diagrams like $m_b$,
are not expected to make an impact. For example, the region of $s'>m_W^2$ in 
Eq.~(\ref{gi}) contributes only about $10^{-4}$ of the coefficients $R_{dd}^{(i)}$ and 
$R_{ds}^{(i)}$.

The factor $r$ is defined as the ratio of the CKM matrix elements,
\begin{eqnarray}
r=\frac{\lambda_{d}}{\lambda_{s}}=\frac{V^*_{cd} V_{ud}}{V^*_{cs} V_{us}}\equiv u+iv,
\label{rc}
\end{eqnarray}
where the real part $u\equiv {\rm Re}(r)$ and the imaginary part $v\equiv {\rm Im}(r)$ have 
been introduced. Equation~(\ref{co1}) contains both the real and imaginary pieces, 
which can be treated separately. The imaginary pieces, simply proportional to 
$v$, do not provide nontrivial constraints. Therefore, we consider the real pieces, 
searching for the values of $u$ and $v$ that minimize the squares of these real pieces 
simultaneously, and then check whether the obtained $u$ and $v$ also diminish the imaginary 
pieces of the conditions. It is equivalent to eliminate the product $V^*_{cd} V_{ud}$ using 
unitarity and to work on the ratio $V^*_{cb}V_{ub}/(V^*_{cs}V_{us})$. We have corroborated that 
this option leads to the same conclusion within theoretical uncertainties.

\begin{figure}
\begin{center}
\includegraphics[scale=0.3]{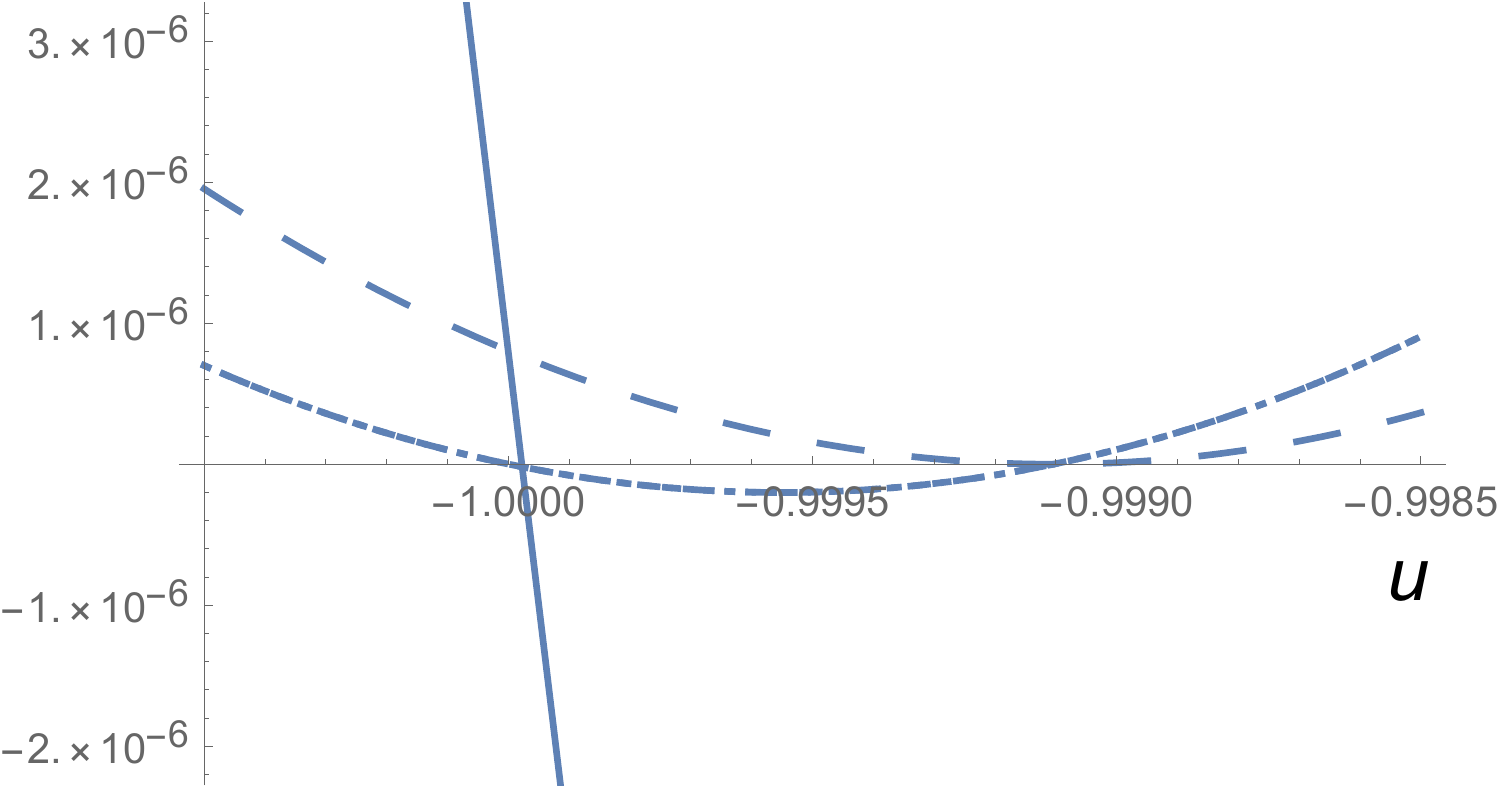}\hspace{1.0cm}
\includegraphics[scale=0.3]{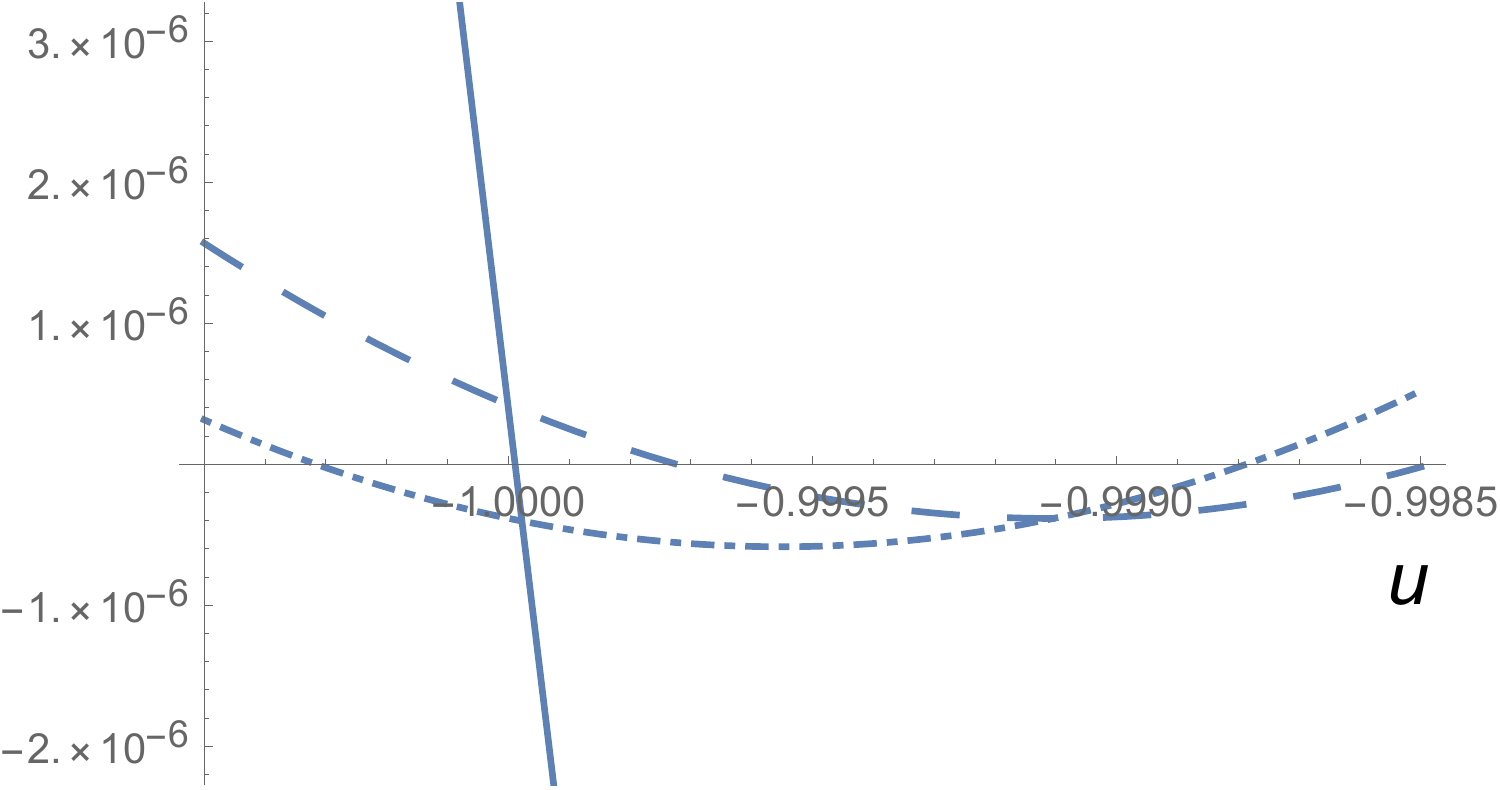}

(a) \hspace{8.0 cm} (b)
\caption{\label{fig1}
(a) Dependences of $(u^2-v^2)R_{dd}^{(m)}+2uR_{ds}^{(m)}+1$ on $u$ with $v=0$ for $m=1$ (dashed line), 
$m=0$ (dotted line), $m=-1$ (dash-dotted line), and $m=i$ (solid line). 
(b) Dependences of $(u^2-v^2)R_{dd}^{(m)}+2uR_{ds}^{(m)}+1$ on $u$ with $v=0.00062$ for $m=1$ 
(dashed line)), $m=-1$ (dash-dotted line), and $m=i$ (solid line).}
\end{center}
\end{figure}

To explain how the aforementioned minima is reached, we exhibit the dependencies of the
real pieces $(u^2-v^2)R_{dd}^{(m)}+2uR_{ds}^{(m)}+1$ on $u$ for $v=0$ in Fig.~\ref{fig1}(a) 
with the inputs of the typical quark masses $m_d=0.005$ GeV, $m_s=0.12$ GeV, and $m_b=4.0$ GeV 
\cite{Li:2023dqi}, and the $W$ boson mass $m_W=80.377$ GeV \cite{PDG}. The definition of 
the quark masses is not very relevant in the current leading-order formalism, and our point 
is to demonstrate that the correct ratio of the CKM matrix elements in Eq.~(\ref{rc}) can be 
produced by the ballpark values of the quark masses. Our results mainly depend on 
the ratios of the quark masses as seen below, such that the renormalization-group 
evolution effects on the quark masses largely cancel. The distinction 
between the curves corresponding to $m=0$ and $m=-1$ is invisible, namely, the two conditions 
are equivalent basically. The two coefficients of the $r^2$ and $r$ terms for $m=0$ 
and $m=-1$ are almost identical up to corrections of $O(m_s^4/(m_W^2m_b^2))\sim 10^{-9}$. As a 
contrast, the coefficients for $m=1$ differ from those for $m=-1$ by $O(m_s^2/m_b^2)\sim 10^{-3}$, 
and from those for $m=i$ by $10^{-2}$. The three curves labeled by $m=1,0,-1$ intersect at the 
location on the horizontal axis, 
\begin{eqnarray}
u=-\frac{(m_W^2-m_d^2)(m_b^2-m_s^2)}{(m_W^2-m_s^2)(m_b^2-m_d^2)},\;\;\;\;v=0,
\end{eqnarray}
where the three conditions $(u^2-v^2)R_{dd}^{(m)}+2uR_{ds}^{(m)}+1=0$ are satisfied exactly. 
The steeper curve from the $m=i$ condition favors the intersection at a more negative 
$u$, and drags the solution to $u\approx -1$. As $v$ increases, the $m=1,0,-1$ curves move downward, 
while the $m=i$ one is relatively stable. 


Below we drop the $m=0$ condition, which is equivalent to the $m=-1$ one, and minimize the sum
\begin{eqnarray}
\sum_{m=1,-1,i}\left[(u^2-v^2)R_{dd}^{(m)}+2uR_{ds}^{(m)}+1\right]^2,\label{sum}
\end{eqnarray}
by varying the unknowns $u$ and $v$. It is legitimate to define this sum, because all terms 
in the above expression have been made dimensionless by taking the ratios. Note that 
$\Gamma_{ij}^{(m)}$ have different dimensions for different $m$, as indicated in Eq.~(\ref{mij}). 
It is easy to find that the above sum decreases with $v$, and arrives at its minimum, when the 
intersection of the $m=i$ curve with the horizontal axis goes between the $m=1$ and $m=-1$ curves
as displayed in Fig.~\ref{fig1}(b). The numerical study coincides with this picture, 
yielding $r=-1.0 +0.00062i$. Simply speaking, the value of $u$ ($v$) is mainly determined by
the $m=i$ condition ($m=1,-1$ conditions). The sign of $v$ cannot be fixed, for this 
unknown appears as $v^2$ in Eq.~(\ref{sum}). We pick up the plus sign for comparison with the 
data, and defer the elaboration on the choice of the opposite sign to the next section.

We estimate the theoretical uncertainties associated with the result of the ratio $r$. 
The subtraction terms in Eq.~(\ref{gi}) just need to cancel the large $s'$ contribution, so 
the lower bounds of their integrations are allowed to vary from $t_{ij}$. We increase the lower 
bounds to $10t_{ij}$ gradually for the three subtraction terms, and observe that the real part 
$u$ is not altered, and the imaginary part $v$ changes by only 3\%, within 0.00060 and 0.00064.
The results of $r$ are insensitive to the $d$ quark mass $m_d$, which can approach zero in fact, 
and depend on $m_s$ and $m_b$ through their ratio $m_s/m_b$. The minimization always returns the 
value $u\approx -1$ with the uncertainty at $10^{-5}$ level under the variation of the quark 
masses, so we scrutinize only the dependence of $v$ on $m_s$. It is found that $v$ increases 
with $m_s$, taking the value $0.00052$ (0.00074) for $m_s=0.11$ (0.13) GeV. Hence, we summarize 
our prediction as
\begin{eqnarray}
r=\frac{V^*_{cd} V_{ud}}{V^*_{cs} V_{us}}=-1.0 +(6.2^{+1.2}_{-1.0})\times 10^{-4}i. \label{rp}
\end{eqnarray}
Inserting the central value of $r$ into the $m=1$ condition, we get
\begin{eqnarray}
r^2R_{dd}^{(1)}+2rR_{ds}^{(1)}+1=(4.04- 10.96i)\times 10^{-7}.\label{go}
\end{eqnarray}
It indicates that the minimization of the real pieces in Eq.~(\ref{co1}) also guarantees
the smallness of the imaginary pieces relative to the constant unity on the left-hand side, 
as claimed before. 

We repeat the dispersive analysis based on the box-diagram contribution associated with 
the $(S-P)(S-P)$ operator \cite{BSS,Cheng},
\begin{eqnarray}
\Gamma^{(S-P)}_{ij}(s)&=&
\frac{1}{s^2}
\frac{\sqrt{s^2-2s(m_i^2+m_j^2)+(m_i^2-m_j^2)^2}}{(m_W^2-m_i^2)(m_W^2-m_j^2)}\nonumber\\
& &\times\left(m_W^4+\frac{m_i^2m_j^2}{4}\right)
[s^2+s(m_i^2+m_j^2)-2(m_i^2-m_j^2)^2],
\label{cij}
\end{eqnarray} 
where the overall coefficient has been also suppressed. The three terms in its asymptotic 
expansion differ from those in Eq.~(\ref{mij}). In particular, the constant term vanishes,
such that the numerical handling in this case is consistent with the ignoring of the 
$m=0$ condition in the $(V-A)(V-A)$ case. The minimization of the sum over the 
$m=1,-1,i$ conditions gives $r=-1.0 +0.00062i$, identical to Eq.~(\ref{rp}) from the $(V-A)(V-A)$ 
contribution. We have to present the values up to three digits in order to reveal the distinction, 
i.e., $v=0.000617$ from $(V-A)(V-A)$ and $v=0.000616$ from $(S-P)(S-P)$. The above examination 
confirms the consistency of our approach.

The CKM matrix is written, in the Chau-Keung (CK) parametrization \cite{CK84}, as
\begin{equation}           
V_{\rm CKM}=\left( \begin{array}{ccc}
    c_{12}c_{13} & s_{12}c_{13} & s_{13}e^{-i\delta}\\
    -s_{12}c_{23}-c_{12}s_{23}s_{13}e^{i\delta} & c_{12}c_{23}-s_{12}s_{23}s_{13}e^{i\delta}&
    s_{23}c_{13}\\
    s_{12}s_{23}-c_{12}c_{23}s_{13}e^{i\delta} & -c_{12}s_{23}-s_{12}c_{23}s_{13}e^{i\delta} &
    c_{23}c_{13}
   \end{array} \right)\;.\label{CKMb}
\end{equation} 
Given the sines of the mixing angles $s_{12}\equiv\sin\theta_{12}=0.22500\pm 0.00067$, 
$s_{13}\equiv\sin\theta_{13}=0.00369\pm 0.00011$, and
$s_{23}\equiv\sin\theta_{23}=0.04182^{+0.00085}_{-0.00074}$, the $CP$ phase  
$\delta=1.144\pm 0.027$ \cite{PDG}, and the corresponding $c_{12}\equiv\cos\theta_{12}$, 
$c_{13}\equiv\cos\theta_{13}$, and $c_{23}\equiv\cos\theta_{23}$, we have the ratio $r$ extracted 
from data with 
\begin{eqnarray}
u=-1.00029\pm 0.00002,\;\;\;\; v=0.00064\pm 0.00002.\label{mea}
\end{eqnarray}
The dominant error of $u$ ($v$) arises from the uncertainty of $\delta$ ($\theta_{13}$).
It is obvious that our determination in Eq.~(\ref{rp}) from the typical quark masses agrees
with the measured values.

Equation~(\ref{go}) implies that the conditions in Eq.~(\ref{co1}) are respected to a good accuracy, 
as the minimum is located. We can then solve for the analytical expression of $v$ by inserting 
$u=-1$ corresponding to the minimum into the $m=1$ condition,
\begin{eqnarray}
v\approx \frac{(m_W^2-m_b^2)(m_s^2-m_d^2)} {(m_W^2-m_s^2)(m_b^2-m_d^2)}\approx
\frac{m_s^2}{m_b^2},\label{vbs}
\end{eqnarray}
where the approximation is valid for the large $m_W$ and small $m_d$. The expansion of the ratio 
$V^*_{cd} V_{ud}/(V^*_{cs} V_{us})$ in the Wolfenstein parameters $\lambda$, $A$, $\rho$ and 
$\eta$ up to $\lambda^4$ \cite{Ahn:2011fg} leads to $v=A^2\lambda^4\eta$.
Equating this expression to Eq.~(\ref{vbs}), we derive the known numerical relation
\begin{eqnarray}
\lambda=V_{us}\approx (A^2\eta)^{-1/4}\sqrt{\frac{m_s}{m_b}}\approx \sqrt{\frac{m_s}{m_b}},
\label{rat}
\end{eqnarray}
with $(A^2\eta)^{-1/4}\approx 1.43\sim O(1)$ for $A\approx 0.826$ and $\eta\approx 0.348$
\cite{PDG}. The above relation manifests the observation that our numerical outcomes of
$r$ mainly depend on the mass ratio $m_s/m_b$.


Another intriguing remark is stimulated by the application of the formalism to the case 
with only two generations of quarks, for which the width difference between the two meson mass 
eigenstates has a simple form
\begin{eqnarray}
\Gamma_{12}(s)\propto \lambda_d^2\left[\Gamma_{dd}(s)-2\Gamma_{ds}(s)+\Gamma_{ss}(s)\right].
\end{eqnarray}
The dispersive analysis on heavy meson lifetimes \cite{Li:2023dqi} has shown that the masses of 
the $d$ and $s$ quarks cannot be degenerate. The dispersion relation in Eq.~(\ref{dis}) with
$M_{12}(s)\approx 0$ at large $s$ is realized only when $\lambda_d$ diminishes, namely, only when
the quark mixing tends to be absent. In other words, there should exist at least three generations 
of fermions in order to facilitate sizable mixing among them in the SM.

\section{NEUTRINO MASS ORDERINGS}

The formulas for the $c\bar u$-$\bar c u$ mixing constructed in the previous 
section apply to the lepton mixing straightforwardly. It is natural to investigate the 
mixing between the $\mu^-e^+$ and $\mu^+e^-$ states, which occurs via the same box diagrams 
but with intermediate neutrino channels. Therefore, the dispersive constraints similar to
those on the quark masses and mixing also appear in the lepton sector. The sensitivity of the 
mixing angles to the mass ratio $m_s/m_b$ hints that it is possible to determine the neutrino 
mass ratio, namely, to discriminate neutrino mass orderings by means of the PMNS matrix elements. 
All the steps follow with the correspondence between the quark masses $m_{d,s,b}$ and the 
neutrino masses $m_{1,2,3}$, and between the ratio in Eq.~(\ref{rc}) and the ratio of the PMNS 
matrix elements $r=U^*_{\mu 1}U_{e1}/(U^*_{\mu 2}U_{e2})$. Here we have assumed that neutrinos 
are of the Dirac type. The condition labeled by $m=0$, 
equivalent to the $m=-1$ condition, can be dropped. Viewing the tiny neutrino masses, we expect 
more serious theoretical uncertainties inherent in the framework, and aim at order-of-magnitude 
estimates. The global fits of data from various groups have produced the consistent parameters 
involved in the PMNS matrix \cite{IES, deSalas:2017kay, Capozzi:2018ubv}. We will illustrate the 
numerical analysis by adopting those obtained in \cite{deSalas:2017kay}.


For the neutrino masses in the NH, we take the mass squared differences
$\Delta m^2_{21} \equiv m^2_{2}-m^2_{1}= (7.55^{+0.20}_{-0.16})\times 10^{-5}$ eV$^2$ 
and $\Delta m^2_{32}\equiv m^2_{3}-m^2_{2}=(2.424\pm 0.03)\times 10^{-3}$ eV$^2$ 
\cite{deSalas:2017kay}. As noticed before, our results are insensitive to the lightest neutrino 
mass, so we choose a small $m_1^2=10^{-6}$ eV$^2$. The values of $m_2^2$ and $m_3^2$ are then 
expressed using $\Delta m^2_{21}$ and $\Delta m^2_{32}$. We minimize the sum of the squared 
deviations in Eq.~(\ref{sum}), deriving the ratio of the PMNS matrix elements 
\begin{eqnarray}
r=\frac{U^*_{\mu 1} U_{e1}}{U^*_{\mu 2} U_{e2}}\approx -1.0-0.02i,\label{pre}
\end{eqnarray}
to which the variations of $\Delta m^2_{21}$ and $\Delta m^2_{32}$ cause only 2\% effects. We 
have selected the minus sign for the value of ${\rm Im}(r)$ as making comparison with the data. 
Viewing the larger ${\rm Im}(r)$, we check the $m=1$ condition in a way similar to Eq.~(\ref{go}), 
and find that it deviates from zero at $10^{-4}$-$10^{-3}$ level. Relative to the constant 
unity on the left-hand side, this deviation is acceptable.

The PMNS matrix is parametrized in the same form as Eq.~(\ref{CKMb}). 
The mixing angles $\theta_{12}= (34.5^{+1.2}_{-1.0})^{\circ}$, 
$\theta_{13}=(8.45^{+0.16}_{-0.14})^{\circ}$, and $\theta_{23}=(47.7^{+1.2}_{-1.7})^{\circ}$, 
and the $CP$ phase $\delta=(218^{+38}_{-27})^{\circ}$ from \cite{deSalas:2017kay} yield the 
measured ratio 
\begin{eqnarray}
r=-(0.738^{+0.050}_{-0.048}) - (0.179^{+0.136}_{-0.125})i, \label{dat}
\end{eqnarray}
where the errors mostly come from the variation of $\delta$. The set of fit parameters 
$\theta_{12}= (33.46^{+0.87}_{-0.88})^{\circ}$, $\theta_{13}=(8.41^{+0.18}_{-0.14})^{\circ}$, 
$\theta_{23}=(47.9^{+1.1}_{-4.0})^{\circ}$, and $\delta=(238^{+41}_{-33})^{\circ}$ from another 
group \cite{Capozzi:2018ubv} leads to $r=-(0.801^{+0.219}_{-0.097}) - (0.265^{+0.090}_{-0.145})i$,
which overlaps with Eq.~(\ref{dat}). The real part ${\rm Re}(r)$ and the lower bound of the 
imaginary part ${\rm Im}(r)$ in Eq.~(\ref{dat}) are in the same order of magnitude
as our prediction in Eq.~(\ref{pre}). That is, the dispersive constraints hold at 
order-of-magnitude level in the NH case.


We employ $\Delta m^2_{21} =  (7.55^{+0.20}_{-0.16})\times 10^{-5}$ eV$^2$ and 
$\Delta m^2_{32}= (-2.50^{+0.04}_{-0.03})\times 10^{-3}$ eV$^2$ for determining the ratio 
$r$ in the IH case \cite{deSalas:2017kay}. The mass of the lightest 
neutrino is set to $m_3^2=10^{-6}$ eV$^2$, and $m_1^2$ and $m_2^2$ are retrieved from 
$\Delta m^2_{21}$ and $\Delta m^2_{32}$ accordingly. We predict
\begin{eqnarray}
r\approx -1.0- O(10^{-5}) i,\label{pr2}
\end{eqnarray}
whose real part is stable against the variations of the mass squared differences. The diminishing 
imaginary part, always maintaining below $10^{-5}$, differs dramatically from the observed ratio
\begin{eqnarray}
r=-(1.03^{+0.05}_{-0.16} )- (0.356^{+0.015}_{-0.048})i,\label{da2}
\end{eqnarray}
inferred by $\theta_{12}=(34.5^{+1.2}_{-1.0})^{\circ}$, 
$\theta_{13}=(8.53^{+0.14}_{-0.15})^{\circ}$, $\theta_{13}=(47.9^{+1.0}_{-1.7})^{\circ}$, and 
$\delta=(281^{+23}_{-27})^{\circ}$ \cite{deSalas:2017kay}. The variations of all fit 
parameters contribute some portions of the errors in Eq.~(\ref{da2}). One can make plots 
similar to Fig.~\ref{fig1}, which display the dependence of each term in Eq.~(\ref{sum}) on 
the real part ${\rm Re}(r)\equiv u$. It is observed, for the IH, that the 
$m=i$ curve becomes close to the $m=1$ one. Namely, we need not increase the imaginary 
part ${\rm Im}(r)\equiv v$ to minimize Eq.~(\ref{sum}); it has reached the minimum with 
${\rm Im}(r)\approx 0$ already.

We conclude that the IH spectrum and 
the corresponding PMNS matrix elements do not obey the dispersive constraints because of the 
apparent disagreement of ${\rm Im}(r)$ between Eqs.~(\ref{pr2}) and (\ref{da2}) even after
the experimental errors are considered. The conclusion should be robust, for the 
inclusion of subleading electroweak corrections to the box diagrams is unlikely to change 
the order of magnitude of our prediction in Eq.~(\ref{pr2}). Indeed, the 
inverted ordering is disfavored by larger $\Delta\chi^2$ of global fits as stated in \cite{PDG}. 
Nevertheless, the closeness of the measured ratios for the NH and IH indicates that it is still 
challenging to discriminate these two spectra experimentally. It is thus encouraging that such 
discrimination can be achieved theoretically in our formalism. Since the extraction of the $CP$ 
phase $\delta$ is more sensitive to the neutrino mass orderings, $\delta\sim 220^\circ$ from the 
NH vs $\delta\sim 280^\circ$ from the IH \cite{PDG}, our observation also helps pin down the 
value of $\delta$. 


We then test the consistency of the QD spectrum under the dispersive
constraints. Taking into account the bound on the neutrino mass sum $\sum m_\nu < 0.12$ eV 
\cite{PDG} at order of magnitude, we assign a sizable value $m_1^2=0.01$ eV$^2$ arbitrarily, and 
write $m_2^2$ and $m_3^2$ by means of $\Delta m^2_{21}$ and $\Delta m^2_{32}$ in the NH.
Other choices of large $m_1^2$ give the same conclusion. The minimization of the sum over the 
squared deviations returns the ratio 
\begin{eqnarray}
r\approx -0.97-O(10^{-5}) i,
\end{eqnarray} 
whose tiny imaginary part does not fit the general feature of the measured PMNS matrix elements 
with ${\rm Im}(r)\sim 10^{-2}$-$10^{-1}$. The above result can be also visualized through plots 
similar to Fig.~\ref{fig1}. Since only the NH scenario passes our dispersive constraints,
we examine the influence from increasing the lowest mass $m_1$ in the NH. It 
reduces ${\rm Im}(r)$ as expected, because a larger $m_1^2$ makes the 
NH ordering closer to the QD spectrum. As $m_1^2$ reaches $1.4\times 10^{-5}$ eV$^2$,
${\rm Im}(r)$ is lowered to $10^{-3}$, which differs from the observed value significantly.
This $m_1^2$ for the NH, together with $\Delta m^2_{21}$ and $\Delta m^2_{32}$, sets an upper 
bound of the neutrino mass sum 
\begin{eqnarray}
\sum m_\nu < 0.082\;\;{\rm eV},
\end{eqnarray} 
which is a bit tighter than the bound in \cite{PDG}.

We are ready to elucidate the different mixing patterns between the quark and lepton sectors
with the solutions at hand. The real parts of $r$ in Eqs.~(\ref{pre}) and
(\ref{dat}) do not differ from $-1$ much, so we insert ${\rm Re}(r)=-1$ into the $m=1$ 
condition, solving for the approximate expression of ${\rm Im}(r)$, which is the same as 
Eq.~(\ref{vbs}) but with the replacement of $m_s$ ($m_b$) by $m_2$ ($m_3$). It is trivial to 
get, from the CK parametrization in Eq.~(\ref{CKMb}) which applies to both the CKM and PMNS 
matrices,
\begin{eqnarray}
{\rm Im}(r)\propto \frac{s_{13}s_{23}}{s_{12}}.\label{ir}
\end{eqnarray}
Here only the sines of the mixing angles 
are highlighted. It is clear that the much larger $\theta_{13}$ and $\theta_{23}$ in the lepton 
sector than in the quark sector trace back to the inequality of the mass ratios,
\begin{eqnarray}
\frac{m_2^2}{m_3^2}\approx 3.1\times 10^{-2} \gg \frac{m_s^2}{m_b^2}\approx 9.0\times 10^{-4},
\end{eqnarray}
where $m_2^2/m_3^2$ is evaluated according to the NH spectrum.


At last, we discuss the dispersive constraints originating from the mixing between the $\tau^-e^+$ 
and $\tau^+e^-$ states, which corresponds to the $t\bar u$-$\bar tu$ mixing in the quark sector. 
It is apparent that the conditions the fermion masses and mixing parameters have to meet
are exactly the same as in the $\mu^-e^+$-$\mu^+e^-$ or $c\bar u$-$\bar cu$ mixing owing to the 
identical intermediate channels in the box diagrams. We remind that the $W$-boson thresholds
should be included in the $t\bar u$-$\bar tu$ mixing, which, however, do not modify the 
following argument. There are only two possible nontrivial outcomes other than those presented in 
the previous sections. First, the products of the mixing matrix elements $\lambda_i\lambda_j$ must 
be small, such that the conditions in Eqs.~(\ref{cons}) and (\ref{gij}) hold automatically. This 
is the case happening to the quark sector with the small mixing angles: for instance, we have 
$|V^*_{ts}V_{us}|^2=A^2\lambda^6\approx 9\times 10^{-5}$ with the Wolfenstein parameter
$\lambda\approx 0.225$, lower than $|V^*_{cs}V_{us}|^2=\lambda^2\approx 5\times 10^{-2}$ by three 
orders of magnitude. Second, the minimization for the same conditions selects ${\rm Im}(r)$ of the 
opposite sign, which, as elaborated shortly, occurs to the lepton sector with the large mixing 
angles. The observed ratios 
$U^*_{\tau 1}U_{e1}/(U^*_{\tau 2}U_{e2})=-(1.231^{+0.078}_{-0.186})+(0.204^{+0.085}_{-0.138})i$ 
from \cite{deSalas:2017kay} and $-(1.139^{+0.139}_{-0.207})+(0.266^{+0.050}_{-0.124})i$ 
from \cite{Capozzi:2018ubv} conform to our
postulation approximately, as they are compared with the corresponding ratios
$U^*_{\mu 1}U_{e1}/(U^*_{\mu 2}U_{e2})$ in and below Eq.~(\ref{dat}).

To understand how the two ratios of the PMNS matrix elements are correlated, we inspect
their explicit expressions in the CK parametrization
\begin{eqnarray}
\frac{U^*_{\mu 1} U_{e1}}{U^*_{\mu 2} U_{e2}}&=&
-\frac{c_{12}}{s_{12}}\frac{c_{12}s_{12}(c_{23}^2-s_{13}^2s_{23}^2)
+c_{23}s_{13}s_{23}c_\delta(c_{12}^2-s_{12}^2)
-c_{23}s_{13}s_{23}s_\delta i}
{(c_{12}c_{23}-s_{12}s_{13}s_{23})^2+2c_{12}c_{23}s_{12}s_{13}s_{23}(1-c_\delta)},\label{ra11}\\
\frac{U^*_{\tau 1} U_{e1}}{U^*_{\tau 2} U_{e2}}&=&
-\frac{c_{12}}{s_{12}}\frac{c_{12}s_{12}(s_{23}^2-c_{23}^2s_{13}^2)
- c_{23}s_{13}s_{23}c_\delta(c_{12}^2-s_{12}^2)
+c_{23}s_{13}s_{23}s_\delta i}
{(c_{12}s_{23}+c_{23}s_{12}s_{13})^2-2c_{12}c_{23}s_{12}s_{13}s_{23}(1-c_\delta)},
\label{ra2}
\end{eqnarray}
with $c_\delta\equiv\cos\delta$ and $s_\delta\equiv\sin\delta$. Note that the imaginary part
of Eq.~(\ref{ra11}) is a complete expression of Eq.~(\ref{ir}). Our solution that 
Eqs.~(\ref{ra11}) and (\ref{ra2}) differ only by the signs of their imaginary parts 
necessitates the rough equalities of the denominators and of the real pieces in the 
numerators, which yield
\begin{eqnarray}
& &(c_{12}^2-s_{12}^2s_{13}^2)(c_{23}^2-s_{23}^2)-4c_{12}c_{23}s_{12}s_{13}s_{23}c_\delta\approx 0,
\label{e1}\\
& &c_{12}s_{12}(1+s_{13}^2)(c_{23}^2-s_{23}^2)+2(c_{12}^2-s_{12}^2)c_{23}s_{13}s_{23}c_\delta
\approx 0,\label{e2}
\end{eqnarray}
respectively. The combination of the above two relations, resulting in
$(c_{12}^2+s_{12}^2s_{13}^2)(c_{23}^2-s_{23}^2)\approx 0$, thus
requires $c_{23}\approx s_{23}$, i.e., $\theta_{23}\approx 45^\circ$ in accordance with the 
observed $\theta_{23}$ around the maximal mixing. The $\mu$-$\tau$ reflection symmetry 
\cite{Harrison:2002et} is thus realized trivially. It is also seen that both Eqs.~(\ref{e1}) 
and (\ref{e2}) can be fulfilled by small $s_{13}$ with $\theta_{23}\approx 45^\circ$.
We mention that new physics effects can deviate $\theta_{23}$ from $45^\circ$. For 
instance, the inclusion of a sequential fourth generation in the similar analysis, 
i.e., the extension of the PMNS matrix to a $4\times 4$ one gives rise to 
$\theta_{23}\approx 47^\circ$ in the second octant \cite{Li:2024xnl}. 




\section{CONCLUSION}

We have deduced the constraints on the fermion masses and mixing parameters from the dispersion 
relations obeyed by the box-diagram contributions to the mixing of two neutral states. These
dispersion relations connect the behaviors of neutral state mixing before and after the electroweak 
symmetry breaking. They are solved with the inputs from the disappearance of the mixing phenomenon 
at high energy, where the electroweak symmetry is restored. The establishment of the solutions 
demands several conditions, which the fermion masses and mixing parameters at low energy, i.e., 
in the symmetry broken phase must satisfy. Taking the $D$ meson, i.e., $c\bar u$-$\bar cu$ mixing as 
an example, we have demonstrated that the typical $d$, $s$ and $b$ quark masses involved in the 
box diagrams specify the ratio of the CKM matrix elements $V^*_{cd} V_{ud}/(V^*_{cs} V_{us})$ in 
agreement with the measured value. Moreover, the imaginary part of the above ratio, as solved 
analytically, generates the known numerical relation $V_{us}\approx \sqrt{m_s/m_b}$. These results 
provide a convincing support to our formalism, which can be refined by including subleading 
corrections to the box diagrams systematically. The constraints obtained in the present paper are 
expected to be modified by these subleading corrections, which may improve the consistency 
between the current predictions and the data, or lead to more precise predictions.

Repeating the same analysis on the $\mu^- e^+$-$\mu^+ e^-$ mixing and the $\tau^- e^+$-$\tau^+ e^-$ 
mixing, which take place via the box diagrams with intermediate neutrino 
channels, we have shown that the neutrino masses in the NH match the observed PMNS matrix 
elements at order-of-magnitude level. The orderings in the IH and QD spectra generate the imaginary 
parts of the ratio $U^*_{\mu 1}U_{e1}/(U^*_{\mu 2}U_{e2})$, which are unequivocally too low 
compared with those from global fits of the data. The leptonic $CP$ phase $\delta$ is
then likely to be in the third quadrant in favor of the NH scenario. The analytical solution 
for the imaginary part of $U^*_{\mu 1}U_{e1}/(U^*_{\mu 2}U_{e2})$ explains the larger lepton 
mixing angles relative to the quark ones via the inequality $m_2^2/m_3^2\gg m_s^2/m_b^2$ 
for $m_{2,3}$ in the NH. Our observation that the ratios
$U^*_{\mu 1}U_{e1}/(U^*_{\mu 2}U_{e2})$ and $U^*_{\tau 1}U_{e1}/(U^*_{\tau 2}U_{e2})$ 
differ only by the sign of their imaginary parts requests the maximal mixing 
$\theta_{23}\approx 45^\circ$, i.e., the $\mu$-$\tau$ reflection symmetry. The above 
summarize the implications from our dispersive analysis on those unresolved issues in neutrino 
physics. 

Combining the previous works, we conjecture that part of the flavor structures in the 
SM, such as the particle masses from 0.1 GeV up to the electroweak scale and the distinct quark 
and lepton mixing patterns, are understood through the internal consistency of SM dynamics.
In other words, the scalar sector of the SM may not be completely free, but 
arranged properly to achieve the dynamical consistency. To maintain this attractive feature, 
a natural extension of the SM is to include
the sequential fourth generation of fermions, since the associated parameters in 
the scalar sector, i.e., their masses and mixing with lighter generations 
can be predicted unambiguously in our formalism. The predictions for the masses 
$m_{b'}\approx 2.7$ TeV and  $m_{t'}\approx 200$ TeV of the sequential 
fourth generation quarks $b'$ and $t'$, respectively, are referred to Ref.~\cite{Li:2023fim}.
We believe that this research direction is worth of further exploration.

\section*{Acknowledgement}

We thank Y.T. Chien, T.W. Chiu, A. Fedynitch, B.L. Hu, Y.H. Lin, M.R. Wu, and T.C. Yuan 
for fruitful discussions. 
This work was supported in part by National Science and Technology Council of the Republic of 
China under Grant No. MOST-110-2112-M-001-026-MY3.


\end{document}